\documentclass[
    ,final            
  ]
  {aipproc}

\layoutstyle{6x9}

\begin{document}

\title{Nuclear Structure and Response based on \\  
  Correlated Realistic NN Interactions
}

\author{P.~Papakonstantinou}{
  address={Institut f\"ur Kernphysik, Technische Universit\"at Darmstadt, 
  \\ Schlossgartenstr.~9, D-64289 Darmstadt, Germany}
}

\author{R.~Roth}{
  address={Institut f\"ur Kernphysik, Technische Universit\"at Darmstadt, 
  \\ Schlossgartenstr.~9, D-64289 Darmstadt, Germany}
}

\author{H.~Hergert}{
  address={Institut f\"ur Kernphysik, Technische Universit\"at Darmstadt, 
  \\ Schlossgartenstr.~9, D-64289 Darmstadt, Germany}
}

\author{N.~Paar}{
  address={Institut f\"ur Kernphysik, Technische Universit\"at Darmstadt, 
  \\ Schlossgartenstr.~9, D-64289 Darmstadt, Germany}
}

\begin{abstract}

 Starting from the Argonne V18 nucleon-nucleon interaction and 
 using the Unitary Correlation Operator Method,  
 a correlated interaction 
 $\mathrm{v}_{\mathrm {UCOM}}$ 
 has been constructed, which is  
 suitable for calculations 
 within restricted Hilbert spaces. 
 In this work 
 we employ the $\mathrm{v}_{\mathrm {UCOM}}$ 
 in Hartree-Fock, perturbation-theory and RPA 
 calculations 
 and we study the ground-state properties 
 of various closed-shell nuclei, as well as some excited states. 
 The present calculations provide also 
 important feedback for the optimization of the 
 $\mathrm{v}_{\mathrm {UCOM}}$ 
 and valuable information on its properties. 
 The above scheme 
 offers the prospect of 
 ab initio calculations in nuclei, regardless of their 
 mass number. 
 It can be used in conjunction 
 with other realistic NN interactions as well, 
 and with various many-body 
 methods (Second RPA, QRPA, Shell Model, etc.). 

\end{abstract}

\maketitle


The Unitary Correlation Operator Method (UCOM)  
provides a novel scheme for carrying out 
nuclear structure calculations 
starting from realistic nucleon-nucleon (NN) 
interactions~\cite{FNR98,NeF03,RNH04,RHP05}. 
The major short-range correlations, 
induced by the strong repulsive core and the tensor part 
of the NN potential, are described by a 
state-independent 
unitary correlation operator $C$. 
This can be used to introduce 
correlations into an uncorrelated $A-$body state or, alternatively, 
to perform a similarity tranformation of an  
operator of interest.  
Applied to a realistic NN interaction, in particular, the method 
produces a "correlated" interaction, 
$\mathrm{v}_{\mathrm {UCOM}}$, 
which can be used 
as a universal effective interaction, 
for calculations 
within simple Hilbert spaces. 
The same transformation can then be applied to any 
other operator under study, as is needed for a consistent UCOM treatment.

The utilization of the UCOM involves a cluster expansion 
of the correlated operators and, currently, 
a truncation at the 
two-body (2B) level. The latter is justified by the 
short range of the correlations treated by the method. 
The aim is to treat explicitly only the state-independent 
short-range correlations (SRC); long-range correlations (LRC) should be 
described by the model space. 
The correlation functions are determined for each $(S,T)$ channel 
by minimizing the energy of the two-nucleon system. 
Three-body interactions are currently not included in the 
formulation; one way to account for these is by adding a 
simple phenomenological non-local 2B correction to the correlated Hamiltonian. 
The introduction of such a correction 
(the same for all nuclei) 
has allowed the UCOM 
to successfully describe properties 
of nuclei up to mass numbers 
$A\approx 60$, 
in the framework of variational calculations 
within Fermionic Molecular Dynamics~\cite{RNH04}. 

In this work we study nuclear structure and response,  
based on realistic interactions, 
without restricting ourselves to light-to-medium systems. 
This is made possible by employing 
the $\mathrm{v}_{\mathrm {UCOM}}$ 
in Hartree-Fock~(HF)- and RPA-based 
models. 
The 2B matrix elements 
of the ${\mathrm v}_{\rm UCOM}$ 
(without correction terms), 
Coulomb interaction and (intrinsic) kinetic energy 
are calculated in a 
harmonic-oscillator (HO) basis.   
These are used as input to subsequent 
HF and RPA calculations, 
in configuration space. 
Only spherical, closed-shell nuclei have been 
considered so far. 
The following methods have been used: 
\begin{description} 
\item[{\rm HF}] - a spherical-HF method, to estimate at "zeroth order" 
the nuclear ground state (gs) properties. 
The HF single-particle states are expanded 
in the HO basis.  
\item[{\rm HF+PT}] 
- second order perturbation theory (PT) is performed on the HF basis to 
obtain a correction to the gs energy.  
\item[{\rm HF+RPA}] - a self-consistent model, which allows us 
to estimate a correction to the gs energy due to LRC, and 
in addition to study collective excitations. 
\item[{\rm ERPA}] - an extended RPA~\cite{CGP98}, which is built on top of the 
true RPA gs and involves an iterative solution of the RPA 
equations. 
Corrected single-particle energies and occupation numbers 
can be obtained, as well as excitation properties.  
\end{description} 
The PT and (E)RPA allow us to account for LRC, 
in addition to the SRC 
introduced via the UCOM. 
The optimal separation of the two types 
of correlations remains an important task. 
It is expressed primarily by a 
constraint   
on the range of the tensor correlation functions, 
imposed during their parameterization. 
By varying this range 
(more precisely, the "correlation volume" $I_{\vartheta}$~\cite{RHP05}), 
a family of correlators and respective correlated 
interactions are obtained. 
Our present results 
provide important feedback 
for the optimization of the 
${\mathrm v}_{\mathrm{UCOM}}$ 
and valuable information on its properties.  

The maximum HO-energy and angular-momentum quantum numbers used 
here are $N_{\max}=2n+\ell=12$ and $\ell_{\max}=10$, 
respectively, providing a satisfactory degree of convergence. 
We use the $\mathrm{v}_{\mathrm{UCOM}}$ parameterizations 
discussed in Ref.~\cite{RHP05}, based on the Argonne V18 
interaction.  
The results presented in Figs.~1 and 2 
were obtained with 
$I_{\vartheta}^{(S=1,T=0)}=0.09$~fm$^3$. 
This value best reproduces, 
in exact no-core shell model calculations, 
the binding energy of the light systems 
$^{4}$He, $^3$H \cite{RHP05}. 

\begin{figure}
  \includegraphics[height=.13\textheight,width=0.70\textwidth]{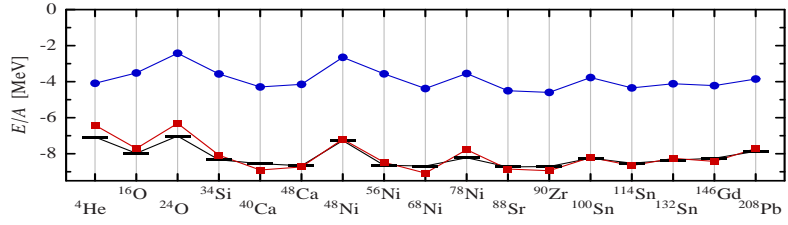}
  \caption{Binding energy per nucleon, for the indicated nuclides, 
in HF (blue dots) and HF+PT (red squares) and experimental (bars). }
\end{figure}
\begin{figure}
  \includegraphics[angle=270,width=0.75\textwidth]{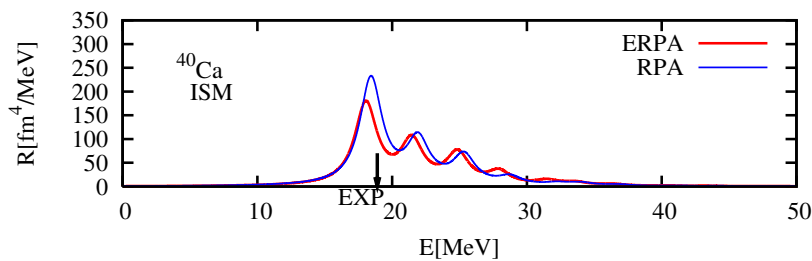}
  \caption{Isoscalar monopole resonance of $^{40}${Ca}, 
in HF+RPA (blue lines) and ERPA (red lines). 
An arrow indicates the experimental centroid.}
\end{figure}

\paragraph{Ground state properties} 
Binding is achieved already at the HF level. The inclusion of LRC 
corrections to the gs energy via PT brings the gs energy very 
close to the experimental data, as shown in Fig.~1. 
The RPA correlations 
result in overbinding~\cite{PPH05}, mainly due to the double-counting of second-order 
corrections inherent in the model. 
In general, 
larger-$I_{\vartheta}$ tensor correlators 
provide stronger binding both at the HF and the RPA level. 
The HF 
single-particle levels (not shown) are too sparse~\cite{PPH05}. 
The omission of a three-body interaction 
and of LRC are responsible for this effect. 
Occupation numbers $n_i=\langle a^{\dagger}_i a_i\rangle$ 
(in standard notation) have been calculated within 
ERPA for $^{16}$O and $^{40}$Ca. 
In principle, correlated operators $C^{\dagger} a^{\dagger}_i a_i C$ 
should be used. 
The small depletion of the Fermi sea that we obtain 
using uncorrelated operators reflects the effect of the gs LRC~\cite{AHP93}. 
 
\paragraph{Collective excitations} 
Our RPA results on the isoscalar (IS) giant monopole resonance (ISGMR), 
for various medium and heavy nuclei, are in good agreement 
with the experimental data. 
An example is shown in Fig.~2, where we see also that  
the gs correlations taken into account in 
ERPA have a relatively small effect. 
In general, 
lower $I_{\vartheta}$ values result in lower ISGMR energies \cite{PPH05}. 
The isovector (IV) dipole strength (not shown) is distributed at energies which are 
too high compared with experiment; 
ERPA is not able to correct for this result. 
{Inclusion of $2p2h$ configurations within Second RPA is expected to 
bring the IVD strength to lower energies.}   

%

\bigskip

In conclusion, the performance of the $\mathrm{v}_{\mathrm{UCOM}}$ 
is very encouraging. Certain aspects of the model 
(e.g., optimal tensor correlators, three-body effects) 
are still open to improvement. 
 The above scheme can be used in conjunction 
 with other realistic (local or non-local) NN interactions, as well as with 
 various many-body 
 methods (Second RPA, QRPA, Shell Model, etc), 
and offers the prospect of ab initio calculations 
across the nuclear chart. 


\bigskip
{ 
Work supported by the Deutsche Forschungsgemeinschaft through contract SFB 634.
} 

\end{document}